\documentclass[11pt]{article}
\usepackage{moriond,epsfig}

\def\be{\begin{equation}}
\def\ee{\end{equation}}
\def\bea{\begin{eqnarray}}
\def\eea{\end{eqnarray}}

\begin{document}
\vspace*{4cm}
\title{HIGH ENERGY NEUTRINO FLUXES FROM COSMIC ACCELERATORS}

\author{C. HETTLAGE\\K. MANNHEIM}

\address{University Observatory G\"{o}ttingen, Geismarlandstra\ss e 11,\\
D-37083 G\"{o}ttingen, Germany}

\maketitle
\abstracts{We constrain high-energy neutrino fluxes with the observed
cosmic ray and gamma ray fluxes, include flavor oscillations and
propagation through Earth, and show that 
blazars could possibly be detected by cubic-kilometer neutrino telescopes.}

\section{Introduction}
As several cubic-kilometer neutrino telescopes are being planned,~\cite{CKT} it is
instructive to ask about the maximum event rate one should expect for these
telescopes. Evidently any answer to this questions depends on some crucial assumptions,
especially concerning the nature of possible neutrino sources.

Following the discussion in Mannheim \emph{et al}.,~\cite{MPR01} we derive an upper flux limit under
the assumptions that (i)
protons do not escape from the
acceleration region (whereas neutral particles do), 
(ii) the power index of the target flux density spectrum (i.e. the synchrotron
photons which accelerated electrons and protons can interact with
photohadronically) is given by $\alpha=-1$, and
(iii) the cosmic ray proton proton flux has the value
\begin{equation}
N_{p,\rm{ obs}}(E_{p})=0.8\times(E_{p}/1\ {\rm GeV})^{-2.75}\ {\rm
cm^{-2}s^{-1}sr^{-1}GeV^{-1}}.
\label{eq:obsCRflux}
\end{equation}
Eq.~\ref{eq:obsCRflux} is an upper limit consistent
with all observations for proton energies $E_{p}$ between $3\times10^{6}$ and
$10^{12}\ {\rm GeV}$.
Assumption (ii) implies that the bolometric gamma-ray luminosity
$L_{\gamma}$ of the sources considered equals twice the neutrino luminosity
$L_{\nu}$:~\cite{RM98}
\[L_{\gamma}=2L_{\nu}.\]
Using particle physics, one may obtain the muon neutrino production rate from the
corresponding neutron production rate:~\cite{RM98}
\begin{equation}
Q_{\nu_{\mu}}(E)\approx83.3Q_{n}(25E).
\label{eq:Qnumu}
\end{equation}
Here and in the following particles and antiparticles are not
distinguished. The cosmic ray proton flux (which is due to the decay of
neutrons leaving the source) is given by 
\begin{equation}
Q_{cr}(E_{p})\approx Q_{n}(E_{p})\times P_{{\rm esc},n}(E_{p}),
\label{eq:Qcr}
\end{equation}
where $P_{{\rm esc},n}$ denotes the neutron escape probability.
 
\section{Constructing neutrino flux bounds}\label{sec:abstract}
The recipe for constructing a generic upper bound on the neutrino flux
can now be stated as follows: Given some neutron production rate one may
compute the corresponding production rates of muon neutrinos and cosmic ray
protons by means of Eqs.~\ref{eq:Qnumu} and~\ref{eq:Qcr}. In order to compute
the neutrino and proton fluxes $I$ at Earth resulting from these production rates, one has
to integrate over all sources in the universe, bearing in mind that
interactions during the flight (e.g. Bethe-Heitler processes and pion
production) change the particle spectra:
\begin{equation}
I(E)\propto\frac{1}{4\pi}\int_{z_{\min}}^{z_{\max}}M(E,z)\frac{(1+z)^{2}}{4\pi
d_{L}^{2}}\frac{dV_{\rm c}}{dz}\frac{dP_{\rm source}}{dV_{\rm
c}}Q((1+z)E,z)dz.
\label{eq:allsources}
\end{equation}
Here $z$ denotes the redshift factor, $V_{\rm c}$ the comoving volume,
$d_{L}$ the luminosity distance, and $dP_{\rm source}/dV_{\rm c}$ the redshift
distribution. $M(E,z)$ takes into account the modification of the spectrum due
to interactions during the flight. For neutrinos $M(E,z)=1$ holds
valid. 

Finally, the expressions for $I_{\nu_{\mu}}$ and $I_{\rm cr}$ thus obtained
are normalized so that $I_{\rm cr}$ is tangent to the observational cosmic ray
proton flux limit (Eq.~\ref{eq:obsCRflux}). If gamma-rays are overproduced for this
normalization (i.e. if $L_{\gamma}=2L_{\nu}$ is greater than the bolometric
diffuse gamma-ray luminosity~\cite{Sre98}, the proton and
neutrino fluxes are reduced appropriately.

\begin{figure}
\begin{center}\resizebox{\textwidth}{!}{\includegraphics{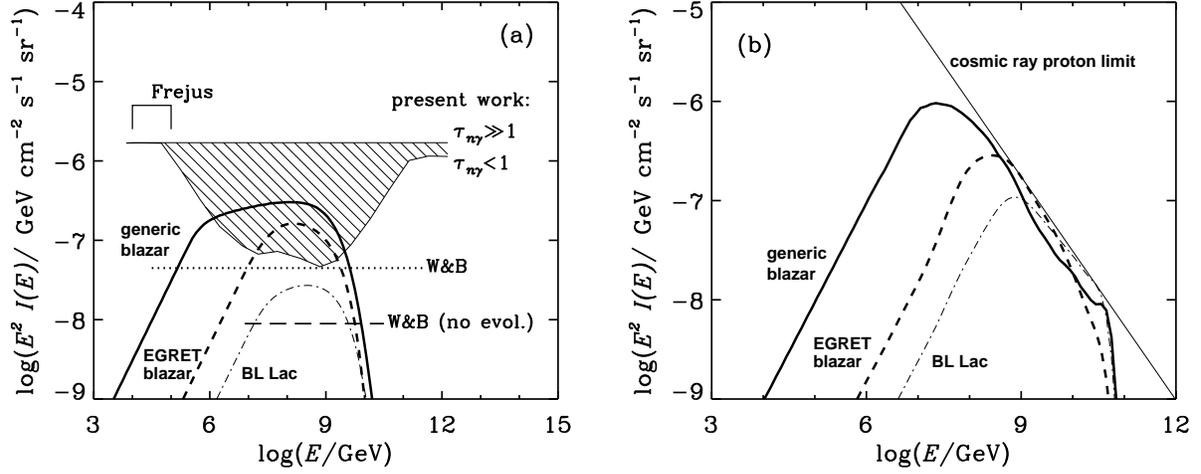}}\end{center}
\caption{\emph{Left:} Cosmic ray bounds on extragalactic neutrino fluxes. The
generic bound for the optically thin ($\tau_{n\gamma}<1$) and thick case
($\tau_{n\gamma}\gg1$), the generic blazar bound (\emph{solid}), the
EGRET blazar
bound (assuming that for $L=10^{48}\ \rm erg/s$ the energy $E_{\tau=1}$ at
which the neutron optical depth is unity has the value
$E_{\tau=1}\approx10^{8}\ \rm GeV$) (\emph{dashed}) and the BL Lac bound
(assuming that $E_{\tau=1}\approx10^{11}\ \rm GeV$ for $L=3\times10^{44}\ \rm erg/s$)
 (\emph{dot-dashed curve}) are shown together
with the Fr\'ejus limit~\protect\cite{Rho96} and the limit inferred by
Waxman and Bahcall (\emph{W}\&\emph{B}) with and without source
evolution.~\protect\cite{WB99} \emph{Right:} Corresponding proton fluxes together with the
cosmic ray proton flux limit used in this work (\emph{straight
line}). Adopted from Mannheim \emph{et~al.}~\protect\cite{MPR01}}
\label{fig:fluxes}
\end{figure}

Fig.~\ref{fig:fluxes} shows several bounds obtained by means of the formalism
just described:
\begin{list}{\arabic{enumi}.}{\usecounter{enumi}\setlength{\parsep}{0.8ex}}
\item Generic bound for optically thin sources with
$Q_{n}(E_{n})\propto E_{n}^{-1}\exp(-E_{n}/E_{\max})$, the redshift distribution
of which equals that of AGN and galaxies.~\cite{BT98} The generic bound
(denoted by $\tau_{n\gamma}<1$) is the
envelope of the flux limits computed for each $E_{\max}$.

\item Generic bound for optically thick sources. Here the bound (denoted by
$\tau_{n\gamma}\gg1$) is derived by demanding consistency with the diffuse
gamma ray background only.

\item Generic bound for blazars. The construction parallels that of the
generic bound for optically thin sources, but we assume a
spectral break between
$10^7$~GeV and $10^{11}$~GeV 
in the escaping cosmic ray flux due to 
opacity effects (still allowing
for $>{\rm GeV}$ photon emission).

\item Bounds for EGRET blazars
and BL Lac objects. In computing the flux limit from
Eq.~\ref{eq:allsources} $Q_{\rm cr}$ and $Q_{\nu}$ are averaged over the
luminosity function and redshift distribution of EGRET-detected blazars
and BL Lacs, respectively.
\end{list}
Details of the calculation of these bounds may be found
in Mannheim~\emph{et~al.}~\cite{MPR01}

\section{Event rates in neutrino telescopes}
So far, we have considered muon neutrinos only. However, as neutrinos are
produced predominantly via pion decays, one should expect the ratio
$Q_{\nu_{e}}:Q_{\nu_{\mu}}:Q_{\nu_{\tau}}\approx1:2:0$ for the various
neutrino production rates. In addition, flavor oscillations during the flight
to Earth lead to a flux ratio of
$I_{\nu_{e}}:I_{\nu_{\mu}}:I_{\nu_{\tau}}=1:1:1$.~\cite{Hus00} Hence in the following, we will
assume equal numbers of neutrinos for the three flavors and change the upper
bounds accordingly.

Furthermore, the neutrino fluxes are altered due to neutrino-nucleon
interactions during the crossing of the inner Earth:~\cite{HM01}
\[\frac{dI_{\nu_{i}}(E)}{dt}=-\sigma_{i,\rm
tot}(E)I_{\nu_{i}}(E)+\sum_{k}\int_{E}^{\infty}dE^\prime
\frac{d\sigma_{k\rightarrow
i}}{dE}(E^{\prime},E)I_{\nu_{k}}(E^{\prime})+\mbox{decay of $\tau$}.\]
In this equation, $i$ and $k$ denote neutrino flavors, $t$ the column number density. $d\sigma_{k\rightarrow
i}(E^{\prime},E)/dE$ constitutes the differential cross section for turning a neutrino of
flavor $k$ and energy $E^{\prime}$ into one of flavor $i$ and energy $E$.
$\sigma_{i,\rm tot}$ is the total (charged and neutral current)
cross section of $\nu_{i}$.

Limiting ourselves to muon neutrinos, we may write the differential muon event rate
$d\dot{N}/dE$ in a neutrino telescope with an effective area $A_{\rm eff}$ as a sum
of two terms covering the muons due to neutrino-nucleon interactions inside
and outside the detector:
\[\frac{d\dot{N}}{dE}=\frac{\rho}{m_{p}}A_{\rm
eff}\left(L\int_{E}^{\infty}dE^{\prime}\frac{d\sigma_{\rm CC}}{dE}(E^{\prime},E)I_{\nu_{\mu}}(E^{\prime})+\int_{0}^{\infty}dx\int_{E}^{\infty}dE^{\prime}\frac{d\sigma_{\rm CC}}{dE}(E^{\prime},E_{0}(E,x))I_{\nu_{\mu}}(E^{\prime})\right).\]
Here $d\sigma_{\rm CC}/dE$ constitutes the differential cross section for
charged current interactions, $\rho$ the density of the detector medium
(i.e. $\rho=1\ \rm g/cm^{3}$ for water), $L$ the detector length, and $m_{p}$
the proton mass. $E_{0}(E,x)$ denotes the energy a muon must
have so that after crossing a distance $x$ of the detector medium (and thus
suffering radiative losses) it retains the energy $E$.

\begin{figure}
\begin{center}\resizebox{\textwidth}{!}{\includegraphics{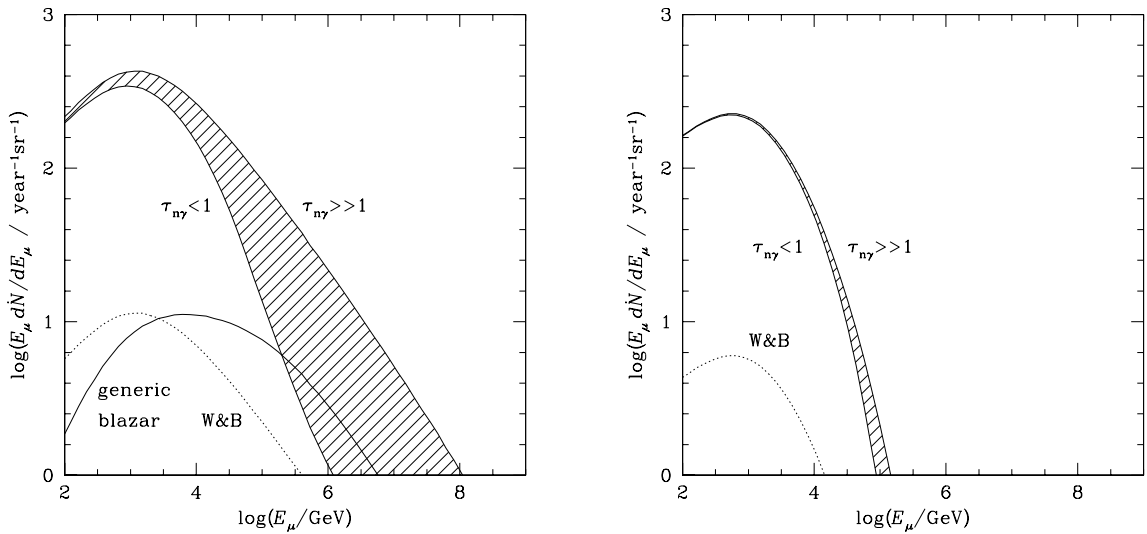}}\end{center}
\caption{\emph{Left:} Muon event rates in a cubic kilometer
\v{C}erenkov telescope, assuming flavor oscillation and horizontal incidence
(no propagation through the Earth). Energy losses outside the detector are
taken into account. The curves correspond to the respective
curves of Fig.~\ref{fig:fluxes}. \emph{Right:} Corresponding rates for
vertical incidence (propagation along the diameter of the  Earth). The generic
blazar event rate lies below $1\ {\rm year^{-1}sr^{-1}}$.} 
\label{fig:events}
\end{figure}

The event rates corresponding to various neutrino flux bounds are shown in
Fig.~\ref{fig:events} for horizontal and vertical incidence, assuming a cubic
kilometer telescope.

\section{Discussion}
Fig.~\ref{fig:events} clearly shows that whereas EGRET-blazar-like 
neutrino spectra might imply an event rate of about ten
events per year and steradian, generic blazar models could allow
neutrino event rates of several hundred events per year and
steradian (most sources would then not be gamma ray emitters
above GeV). This exceeds the limit obtained by Waxman and Bahcall
~\cite{WB99} for optically thin sources (which is comparable to our BL Lac case) by one to two orders of magnitude, reflecting the
different assumptions.

Any derivation of an upper neutrino flux limit faces two important
unknowns:~\cite{MPR01} Firstly, intergalactic magnetic fields (which affect
protons but have no influence on neutrinos) might considerably change the
number of neutrinos compatible with cosmic ray observations. Secondly, the
same applies to exotic processes such as the decay of
superheavy particles.  

The planned cubic-kilometer neutrino telescopes will reach the sensitivity
necessary to probe the assumptions that enter the flux-limit calculations.
If photohadronic processes play an important role in blazars,
we have shown that they will very likely be detected by such telescopes.

\section*{Acknowledgments}
This work was supported by the Studienstiftung des deutschen Volkes. 
We gratefully acknowledge the support of the ESF
Neutrino Network, and we would like to thank the Moriond organizers.

\section*{References}
\bibliography{literature}

\begin{thebibliography}{1}

\bibitem{CKT}
cf. the articles of V. Bertin, G. Riccobene and S. Tzamarias in this volume.

\bibitem{MPR01}
K.~Mannheim, R.~Protheroe, and J.R. Rachen.
\newblock {\em Phys. Rev.}, D63:023003, 2001.

\bibitem{RM98}
J.P. Rachen and P.~M{\'{e}}sz{\'{a}}ros.
\newblock {\em Phys. Rev.}, D58:123005, 1998.

\bibitem{Sre98}
P.~Sreekumar et~al.
\newblock {\em Astrophys. J.}, 494:523, 1998.

\bibitem{BT98}
B.J. Boyle and R.J. Terlevich.
\newblock {\em Mon. Not. R. Astron. Soc.}, 293:L49, 1998.

\bibitem{Rho96}
W.~Rhode et~al.
\newblock {\em Astropart. Phys.}, 4:217, 1996.

\bibitem{WB99}
E.~Waxman and J.~Bahcall.
\newblock {\em Phys.Rev.}, D59:023002, 1999.

\bibitem{Hus00}
A.~Husain.
\newblock {\em Nucl. Phys. B (Proc. Suppl.)}, 87:442, 2000.

\bibitem{HM01}
C.~Hettlage and K.~Mannheim.
\newblock {\em Nucl. Phys. B (Proc. Suppl.)}, 95:165, 2001.

\end{thebibliography}

\end{document}